\documentclass{pasj01}

\usepackage[switch,mathlines]{lineno}

\usepackage[varg]{txfonts}
\usepackage[T1]{fontenc}
\usepackage{natbib,bm}

\usepackage{savesym}
\savesymbol{tablenum}
\usepackage{siunitx}
\restoresymbol{SIX}{tablenum}

\usepackage{mathrsfs}

\begin{document}

\title{Survivability of Amorphous Ice in Comets Depends on the Latent Heat of Crystallization of Impure Water Ice}
\author{Sota \textsc{Arakawa}\altaffilmark{1}}
\altaffiltext{1}{Yokohama Institute for Earth Sciences, Japan Agency for Marine-Earth Science and Technology, Yokohama, Kanagawa, 236-0001, Japan}
\email{arakawas@jamstec.go.jp}

\author{Shigeru \textsc{Wakita}\altaffilmark{2}}
\altaffiltext{2}{Department of Earth, Atmospheric and Planetary Sciences, Massachusetts Institute of Technology, Cambridge, MA, 02139-4307, USA}


\maketitle

\begin{abstract}
Comets would have amorphous ice rather than crystalline one at the epoch of their accretion.
Cometary ice contains some impurities that govern the latent heat of ice crystallization, $L_{\rm cry}$.
However, it is still controversial whether the crystallization process is exothermic or endothermic.
In this study, we perform one-dimensional simulations of the thermal evolution of km-sized comets and investigate the effect of the latent heat.
We find that the depth where amorphous ice can survive significantly depends on the latent heat of ice crystallization.
Assuming the cometary radius of $2~\si{km}$, the depth of the amorphous ice mantle is approximately $100~\si{m}$ when the latent heat is positive (i.e., the exothermic case with $L_{\rm cry} = + 9 \times 10^{4}~\si{J.kg^{-1}}$).
In contrast, when we consider the impure ice representing the endothermic case with $L_{\rm cry} = - 9 \times 10^{4}~\si{J.kg^{-1}}$, the depth of the amorphous ice mantle could exceed $1~\si{km}$.
Although our numerical results indicate that these depths depend on the size and the accretion age of comets, the depth in a comet with the negative latent heat is a few to several times larger than the positive case for a given comet size. 
This work suggests that the spatial distribution of the ice crystallinity in a comet nucleus depends on the latent heat, which can be different from the previous estimates assuming pure water ice.
\end{abstract}


\section{Introduction}

Small icy objects, comets, are composed of ice and refractory materials, and comets may have formed in the outer region of the solar nebula.
The cometary ice is not pure ${\rm H}_{2}{\rm O}$ but have some impurities, such as ${\rm H}_{2}{\rm O}$, ${\rm C}{\rm O}$, ${\rm C}{\rm O}_{2}$, and other substances \citep[e.g.,][]{1988PhRvB..38.7749B, 2010PCCP...12.5947B, 2020SSRv..216..102R}.
The detection of supervolatile gases, including ${\rm N}_{2}$ and ${\rm Ar}$, on comet 67P/Churyumov--Gerasimenko indicates that the temperature of the nucleus was lower than $30~\si{K}$ at the time of its accretion \citep[e.g.,][]{2015Sci...348..232R}.
Original cometary ice is believed to be amorphous rather than crystalline \citep[e.g.,][]{2008SSRv..138..147P, 2022arXiv220905907P}.
\citet{1994A&A...290.1009K} noted that the infrared ${\rm H}_{2}{\rm O}$ ice absorption features observed in molecular clouds are consistent with amorphous ice \citep[see also][]{1985MNRAS.214..289V, 1992dge..book.....W}.  
Based on the timescales of condensation and preservation of amorphous ice, they concluded that amorphous ice would have formed in a parental molecular cloud of the solar system.
\citet{2014ApJ...784L...1C}, however, argued that photodesorption and recondensation of ${\rm H}_{2}{\rm O}$ molecules near the surface layers of the solar nebula could also have formed amorphous ice.

\begin{figure}
\begin{center}
\includegraphics[width=\columnwidth]{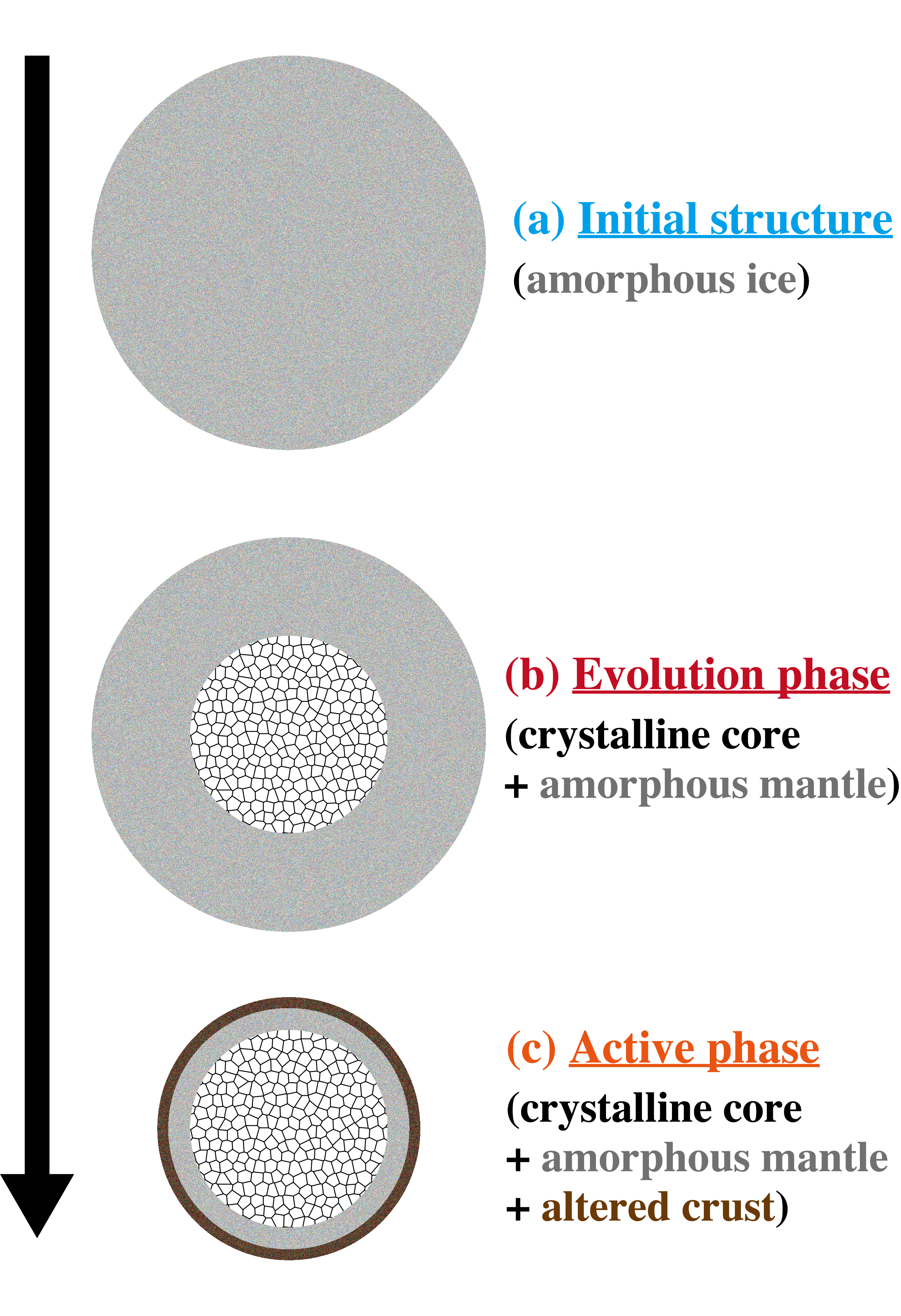}
\end{center}
\caption{
Schematic images of the evolutionary phases of a comet nucleus.
(a) In the accretion phase, comets are formed via the accumulation of ice and refractory dust particles in the gaseous solar nebula.
(b) The interior of comets is thermally evolved due to the decay heat of radioactive nuclides in the evolution phase.
As the temperature increases, the crystalline core is formed. 
(c) In the active phase, intensive cometary activities are driven by the heating due to solar irradiation.
The thermally altered surface becomes a dusty crust, and the mass loss would cause the shrinking of comet nuclei.
}
\label{fig1}
\end{figure}

The evolutionary history of comets can be divided into three distinct phases \citep[e.g.,][]{2004come.book..317M, 2004come.book..359P}: (a) accretion phase, (b) evolution phase, and (c) active phase.
Here we briefly explain the physical state of the comet nucleus in each phase (Figure \ref{fig1}).

In the accretion phase (Figure \ref{fig1}(a)), comets are formed via accumulation of ice and refractory dust particles in the gaseous solar nebula \citep[e.g.,][]{2022arXiv221204509S}.
We assume that original cometary ice is totally amorphous.
Recent laboratory-based studies proposed a hypothesis that the building block of comets is mm- to cm-sized dust aggregates called ``pebbles'' \citep[e.g.,][]{2017MNRAS.469S.755B, 2018SSRv..214...52B}, and the initial structure of comets is deemed ``pebble-pile'' aggregates, i.e., agglomerates of pebbles \citep[e.g.,][]{2012Icar..221....1S, 2020MNRAS.497.1166A, 2022MNRAS.514.3366M, 2022Univ....8..381B}.

In the evolution phase (Figure \ref{fig1}(b)), the interior of comets is thermally evolved due to the decay heat of radioactive nuclides.
The main heat source during this early stage is the short-lived nuclide ${}^{26}{\rm Al}$ whose half-life time is $0.72~\si{Myr}$ \citep[e.g.,][]{1987ApJ...319..993P, 1995Icar..117..420P}.
In addition, earlier studies on the early thermal evolution of comets showed that the latent heat of ice crystallization would play a crucial role \citep[e.g.,][]{1993JGR....9815079H, 2022MNRAS.514.3366M}.
The temperature at the center becomes higher than the surface temperature, and the crystalline core and amorphous mantle would be formed after the early thermal evolution.
Comets would be stored for billions of years in the outermost part of the solar system (i.e., the Oort cloud and the Kuiper belt) before entering the inner solar system where the temperature is sufficiently high to drive cometary activities \citep[e.g.,][]{2015SSRv..197..191D}.

Finally, in the active phase (Figure \ref{fig1}(c)), intensive cometary activities are driven by the heating due to solar irradiation.
There are many potential mechanisms for driving activities, including the sublimation of ices \citep[e.g.,][]{2020MNRAS.493.4039F} and the release of the latent heat of ice crystallization \citep[e.g.,][]{1992A&A...258L...9P}.
The surface of comets is thermally altered and a dusty crust would be formed in the active phase.
In contrast, the heating due to solar irradiation barely affects the inner crystalline core.
The mass loss due to the intensive activities would also cause the shrinking of comet nuclei.

The release of the latent heat of ice crystallization is the pivotal process in the evolution phase.
However, the value (and sign) of the latent heat of crystallization of cometary ice is poorly constrained.
\citet{2001GeoRL..28..827K} reported that the latent heat is drastically reduced by the presence of impurities; while the crystallization of pure ${\rm H}_{2}{\rm O}$ ice is an exothermic process, the crystallization of impure ice is an endothermic process when the concentration of impurities exceeds approximately 2\%.
The molecular abundances in cometary atmospheres are indeed about 1\% for several species including ${\rm C}{\rm O}$, ${\rm C}{\rm O}_{2}$, and ${\rm N}{\rm H}_{3}$ \citep[e.g.,][]{2022arXiv220704800B}.
Therefore, we can expect that the existence of impurities in cometary ice could have a great impact on the latent heat of ice crystallization.

In this study, we perform one-dimensional simulations of the thermal evolution of km-sized comets to investigate the impacts of latent heat of impure ice.
We also explore the depth where amorphous ice can survive would depend on the latent heat of ice crystallization.
The structure of this paper is as follows.
In Section \ref{sec:model}, we introduce the thermal evolution model of spherical comet nucleus.
In Section \ref{sec:results}, we present numerical results for thermal evolution in the evolution phase.
We discuss the possible impacts of thermal evolution on observational features of comets in Section \ref{sec:discussion}, and a summary is presented in Section \ref{sec:conclusion}.

\section{Model}
\label{sec:model}

To understand the thermal evolution of pebble-pile comets, we perform one-dimensional simulations. 
Assuming spherical nucleus, we solve a heat conduction equation considering the release of latent heat due to ice crystallization and the decay heat of radioactive nuclides.
The material properties assumed in this study are listed in Table \ref{table1}.

\begin{table*}
\caption{List of material properties.}
\label{table1}
\centering
\begin{tabular}{lccl}
{\bf Parameter}                                          & {\bf Symbol}       & {\bf Value}                                & {\bf Reference}  \\ \hline
Filling factor inside pebbles                            & $\phi_{\rm peb}$   & $0.4$                                      & \citet{2009ApJ...696.2036W} \\
Filling factor of pebble packing                         & $\phi_{\rm pack}$  & $0.6$                                      & \citet{2014Icar..235..156B} \\
\\
Volume fraction of ice                                   & $\chi_{\rm ice}$   & $0.4$                                      & --- \\
Volume fraction of rock                                  & $\chi_{\rm rock}$  & $0.6$                                      & --- \\  
Material density of ice                                  & $\rho_{\rm ice}$   & $917~\si{kg.m^{-3}}$                       & \citet{2021MNRAS.505.5654D} \\
Material density of rock                                 & $\rho_{\rm rock}$  & $3000~\si{kg.m^{-3}}$                      & \citet{2021MNRAS.505.5654D} \\
\\
Specific heat capacity of ice                            & $c_{\rm ice}$      & Equation (\ref{eq:c_ice})                  & \citet{2022MNRAS.514.3366M} \\
Specific heat capacity of rock                           & $c_{\rm rock}$     & Equation (\ref{eq:c_rock})                 & \citet{2022MNRAS.514.3366M} \\
Thermal conductivity of pebble                           & $k_{\rm peb}$      & Equation (\ref{eq:k_peb})                  & \citet{2023MNRAS.521.4927A} \\
Thermal conductivity of amorphous ice                    & $k_{\rm amo}$      & Equation (\ref{eq:k_mat_amo})              & \citet{2022MNRAS.514.3366M} \\
Thermal conductivity of crystalline ice                  & $k_{\rm cry}$      & Equation (\ref{eq:k_mat_cry})              & \citet{2022MNRAS.514.3366M} \\
\\
Pebble radius                                            & $r_{\rm peb}$      & $1~\si{cm}$                                & --- \\
Particle radius                                          & $r_{\rm par}$      & $0.1~\si{\micro m}$                        & \citet{2007ApJ...661..320W} \\
Surface energy of ice                                    & $\gamma_{\rm par}$ & $100~\si{mJ.m^{-2}}$                       & \citet{2007ApJ...661..320W} \\
Young's modulus of ice                                   & $E_{\rm par}$      & $7~\si{GPa}$                               & \citet{2007ApJ...661..320W} \\
Poisson's ratio of ice                                   & $\nu_{\rm par}$    & $0.25$                                     & \citet{2007ApJ...661..320W} \\
\\
Constant of crystallization rate                         & $A_{\rm cry}$      & $9.54 \times 10^{-14}~\si{s}$              & \citet{1989ESASP.302...65S} \\
Activation energy for crystallization                    & $E_{\rm cry}$      & $7.41 \times 10^{-20}~\si{J}$              & \citet{1989ESASP.302...65S} \\
\\
Decay heat of $^{26}{\rm Al}$ at $t = 0$                 & $H_{26}$           & $1.926 \times 10^{-7}~\si{W.kg^{-1}}$      & \citet{2021MNRAS.505.5654D} \\
Decay heat of $^{40}{\rm K}$ at $t = 0$                  & $H_{40}$           & $2.826 \times 10^{-11}~\si{W.kg^{-1}}$     & \citet{2021MNRAS.505.5654D} \\
Decay heat of $^{232}{\rm Th}$ at $t = 0$                & $H_{232}$          & $1.272 \times 10^{-12}~\si{W.kg^{-1}}$     & \citet{2021MNRAS.505.5654D} \\
Decay heat of $^{235}{\rm U}$ at $t = 0$                 & $H_{235}$          & $4.111 \times 10^{-12}~\si{W.kg^{-1}}$     & \citet{2021MNRAS.505.5654D} \\
Decay heat of $^{238}{\rm U}$ at $t = 0$                 & $H_{238}$          & $1.944 \times 10^{-12}~\si{W.kg^{-1}}$     & \citet{2021MNRAS.505.5654D} \\
Half-life time of $^{26}{\rm Al}$                        & $\tau_{26}$        & $7.05 \times 10^{5}~\si{yr}$               & \citet{2021MNRAS.505.5654D} \\
Half-life time of $^{40}{\rm K}$                         & $\tau_{40}$        & $1.2 \times 10^{9}~\si{yr}$                & \citet{2021MNRAS.505.5654D} \\
Half-life time of $^{232}{\rm Th}$                       & $\tau_{232}$       & $1.4 \times 10^{10}~\si{yr}$               & \citet{2021MNRAS.505.5654D} \\
Half-life time of $^{235}{\rm U}$                        & $\tau_{235}$       & $6.9 \times 10^{8}~\si{yr}$                & \citet{2021MNRAS.505.5654D} \\
Half-life time of $^{238}{\rm U}$                        & $\tau_{238}$       & $4.5 \times 10^{9}~\si{yr}$                & \citet{2021MNRAS.505.5654D} \\
\hline
\end{tabular}
\end{table*}

\subsection{Heat conduction equation}

We adopt thermal evolution models of spherical small bodies \citep[e.g.,][]{2011EP&S...63.1193W, 2014M&PS...49..228W, 2017ApJ...836..106W, 2018ApJ...863..100W}.
We assume that comets are spherically symmetric bodies.
For simplicity, we also assume that comets form at a given time, $t = t_{\rm acc}$, and we disregard their further growth and destruction (i.e., the radius of comet nucleus, $R_{\rm coment}$, is constant over time).
We describe the spatial and temporal variation of the temperature, $T$, within the comet nucleus as a function of the time, $t$, and the distance from the center, $R$, using the heat conduction equation:
\begin{equation}
\rho_{\rm comet} c_{\rm comet} \frac{{\rm d}T}{{\rm d}t} = - \frac{1}{4 \pi R^{2}} \frac{{\partial}F}{{\partial}R} + \dot{Q}_{\rm cryst} + \dot{Q}_{\rm decay},
\label{eq:conduction}
\end{equation}
where $\rho_{\rm comet}$ is the bulk density of comet, $c_{\rm comet}$ is the specific heat capacity, $\dot{Q}_{\rm cryst}$ is the heat generation by crystallization of amorphous ice (see Section \ref{sec:latent}), and $\dot{Q}_{\rm decay}$ is the heat generation by the decay of radioactive nuclides (see Section \ref{sec:decay}). 
The outward energy flux by heat conduction at $R$, $F = F {( R )}$, is given by
\begin{equation}
F = - 4 \pi R^{2} k_{\rm comet} \frac{{\partial}T}{{\partial}R},
\end{equation}
where $k_{\rm comet}$ is the thermal conductivity (see Section \ref{sec:k_and_c}).
The surface temperature at $R = R_{\rm comet}$, $T_{\rm surf}$, is given by the following energy balance equation \citep[e.g.,][]{1993JGR....9815079H}:
\begin{equation}
4 \pi {R_{\rm comet}}^{2} \sigma_{\rm SB} {\left( 1 - A \right)} {\left( {T_{\rm surf}}^{4} - {T_{\rm bg}}^{4} \right)} = F_{\rm surf}, \label{eq:F_surf}
\end{equation}
where $\sigma_{\rm SB}$ is the Stefan--Boltzmann constant, $A = 0.1$ is the surface albedo, $T_{\rm bg} = 40~\si{K}$ is the background temperature (i.e., disk temperature at the orbit of a comet), and $F_{\rm surf}$ is the outward energy flux at the surface (i.e., $F_{\rm surf} = F|_{R \to R_{\rm comet}}$).
We set the comet's uniform initial temperature of $T = T_{\rm bg}$ at $t = t_{\rm acc}$.

When we solve the heat conduction equation (Equation (\ref{eq:conduction})) by the finite-difference method, we use the constant length of the mesh of ${\Delta R} = 2~\si{m}$ regardless of $R_{\rm comet}$.
For the timescale for heat conduction to the next mesh, we consider $t_{\rm mesh}$ as  
\begin{equation}
t_{\rm mesh} = \frac{\rho_{\rm comet} c_{\rm comet}}{2 k_{\rm comet}} {\Delta R}^{2}.
\end{equation}
We describe the temperature dependence of $t_{\rm mesh}$ in Appendix \ref{app:timescale} with Figure \ref{fig:timescale}.

We determine an appropriate value of timestep (${\Delta t}$) that satisfies the following two requirements for all meshes.
The first requirement is ${\Delta t} \ll t_{\rm mesh}$, which represents the Courant--Friedrichs--Lewy condition \citep{1928MatAn.100...32C}.
The second requirement, ${\Delta t} \ll {| {\rm d}C / {\rm d}t |}^{-1}$ (see Section \ref{sec:latent} for details of the crystallinity $C$), allows us to precisely calculate the temporal evolution of the crystallinity.
We set ${\Delta t}$ as the spatial minimum value of $\min{[ 10^{-1} t_{\rm mesh}, 10^{-4} {| {\rm d}C / {\rm d}t |}^{-1} ]}$, and we update ${\Delta t}$ at each time step.\footnote{
Our choice of ${\Delta t}$ would be sufficiently small to precisely calculate the temporal evolution.
We confirmed that the temporal evolution of $C$ (and $T$) is barely changed, even if we set ${\Delta t} = \min{[ 10^{-1} t_{\rm mesh}, 10^{-2} {| {\rm d}C / {\rm d}t |}^{-1} ]}$.
}

We assume that the interior of the comet has uniform density for simplicity.
The bulk density of a pebble-pile comet, $\rho_{\rm comet}$, is given by 
\begin{equation}
\rho_{\rm comet} = \phi_{\rm peb} \phi_{\rm pack} {\left( \chi_{\rm ice} \rho_{\rm ice} + \chi_{\rm rock} \rho_{\rm rock} \right)},
\end{equation}
where $\phi_{\rm peb}$ is the filling factor inside pebbles, $\phi_{\rm pack}$ is the filling factor of pebble packing, $\rho_{\rm ice}$ is the material density of ice, and $\rho_{\rm rock}$ is the material density of rock.
We set that the volume fractions of ice and rock are $\chi_{\rm ice} = 0.4$ and $\chi_{\rm rock} = 0.6$, respectively.
These values are chosen to roughly reproduce the bulk density of comet 67P/Churyumov--Gerasimenko \citep[$532~\si{kg.m^{-3}}$;][]{2016Icar..277..257J}, and it is consistent with that derived for comet 67P using different approaches \citep[e.g.,][]{2019MNRAS.483.2337P, 2019MNRAS.482.3326F}.
The mass fraction of ice, $f_{\rm ice}$, is given by
\begin{equation}
f_{\rm ice} = \frac{\chi_{\rm ice} \rho_{\rm ice}}{\chi_{\rm ice} \rho_{\rm ice} + \chi_{\rm rock} \rho_{\rm rock}},
\end{equation}
and the mass fraction of rock, $f_{\rm rock}$, is
\begin{equation}
f_{\rm rock} = \frac{\chi_{\rm rock} \rho_{\rm rock}}{\chi_{\rm ice} \rho_{\rm ice} + \chi_{\rm rock} \rho_{\rm rock}}.
\end{equation}

\subsection{Latent heat of ice crystallization}
\label{sec:latent}

We investigate the impacts of latent heat of ice crystallization, $L_{\rm cry}$, on the early thermal evolution of km-sized comets. 
Positive or negative values of latent heat reflect the concentration of impurities in cometary ice.
To evaluate the effect of impurities, we regard $L_{\rm cry}$ as a parameter, ranging from $- 9 \times 10^{4}~\si{J.kg^{-1}}$ to $9 \times 10^{4}~\si{J.kg^{-1}}$.
For the pure amorphous ${\rm H}_{2}{\rm O}$ ice, crystallization is an exothermic process (i.e., $L_{\rm cry} > 0$) and $L_{\rm cry} = 9 \times 10^{4}~\si{J.kg^{-1}}$ \citep{1968JChPh..48..503G}.
In contrast, \citet{2001GeoRL..28..827K} reported that $L_{\rm cry}$ decreases when amorphous ${\rm H}_{2}{\rm O}$ ice contains other molecules such as ${\rm C}{\rm O}$, ${\rm C}{\rm O}_{2}$, and ${\rm C}{\rm H}_{4}$. 
When the impurity concentration exceeds 2--3\%, crystallization of amorphous ice is not exothermic but endothermic \citep[i.e., $L_{\rm cry} < 0$;][]{2001GeoRL..28..827K}.

The heat generation by crystallization of amorphous ice per unit volume per unit time, $\dot{Q}_{\rm cryst}$, is given by the following differential equation \citep[e.g.,][]{1993JGR....9815079H}:
\begin{equation}
\dot{Q}_{\rm cryst} = f_{\rm ice} \rho_{\rm comet} L_{\rm cry} \frac{{\rm d}C}{{\rm d}t},
\end{equation}
where $C$ is the crystallinity of ice (i.e., a fraction of crystalline ice).
The temporal evolution of $C$ is given by the following equation, 
\begin{equation}
\frac{{\rm d}C}{{\rm d}t} = \frac{1 - C}{t_{\rm cry} + t_{\rm cry, min}}.
\label{eq:C}
\end{equation}
Here, $t_{\rm cry}$ is the crystallization timescale and it is given as a function of $T$ as follows \citep[e.g.,][]{1989ESASP.302...65S}, 
\begin{equation}
t_{\rm cry} = A_{\rm cry} \exp{\left( \frac{E_{\rm cry}}{k_{\rm B} T} \right)},
\end{equation}
where $A_{\rm cry}$ is the constant of crystallization rate and $E_{\rm cry}$ is the activation energy for crystallization.
In this study, we set $t_{\rm cry, min} = 10^{9}~\si{s}$ to avoid an unphysical increase of the temperature in numerical simulations (see Appendix \ref{app:timescale}).

\subsection{Decay heat of radioactive nuclides}
\label{sec:decay}

We consider the heat generation by the decay of radioactive nuclides per unit volume per unit time, $\dot{Q}_{\rm decay}$. 
If there are several radioactive nuclides within a comet, $\dot{Q}_{\rm decay}$ is given by
\begin{equation}
\dot{Q}_{\rm decay} = f_{\rm rock} \rho_{\rm comet} {\sum_{i} H_{i} \exp{\left( - \frac{t}{\tau_{i} / \ln{2}} \right)}},
\end{equation}
where $H_{i}$ is the decay heat of the radioactive nuclide $i$ by the unit mass of rock at $t = 0$, and $\tau_{i}$ is the half-life time of species $i$.
We consider five species as the source of decay heat; $^{26}{\rm Al}$ ($i = 26$), $^{40}{\rm K}$ ($i = 40$), $^{232}{\rm Th}$ ($i = 232$), $^{235}{\rm U}$ ($i = 235$), and $^{238}{\rm U}$ ($i = 238$).
We take both $H_{i}$ and $\tau_{i}$ from \citet{2021MNRAS.505.5654D}.\footnote{
\citet{2021MNRAS.505.5654D} provided $H_{i}$ in their Table 11.
They calculated it using literature data of elemental mass fractions in CI chondrites \citep{2003ApJ...591.1220L} and isotope molar fractions at $t = 0$ \citep{2012A&A...537A..45H}.
}
In this study, we regard that $t = 0$ is the formation time of calcium--aluminum-rich inclusions (CAIs, 4568 Myr ago).\footnote{
\citet{2021MNRAS.505.5654D} took the canonical ${( ^{26}{\rm Al} / ^{27}{\rm Al} )}_{0}$ value of $5.1 \times 10^{-5}$ from \citet{2009GeCoA..73.5115N}.
}
Although there have been discussions about the heterogeneous distribution of radionuclides \citep[for $^{26}{\rm Al}$, see][]{1995Metic..30..365M, 2002Sci...297.1678A, 2011ApJ...735L..37L, 2013M&PS...48.1383K}, we simply assume that radioactive nuclides are homogeneously distributed within the solar system at $t = t_{\rm acc}$.
We ignore the contribution of $^{60}{\rm Fe}$ in this study.
This is because recent precise measurements of the iron isotope ratio of chondrites \citep[e.g.,][]{2022ApJ...929..107K, 2022ApJ...940...95K} indicated that the initial abundance of $^{60}{\rm Fe}$ would be order(s) of magnitude lower than that assumed in the earlier studies of the thermal evolution of small bodies.

\subsection{Thermal conductivity and specific heat capacity}
\label{sec:k_and_c}

\subsubsection{Thermal conductivity}

The thermal conductivity of a pebble-pile comet, $k_{\rm comet}$, is given by the sum of the radiative and conductive terms: $k_{\rm comet} = k_{\rm rad} + k_{\rm net}$.
The thermal conductivity through the solid particle contacts (i.e., network conduction), $k_{\rm net}$, is previously studied \citep[e.g.,][]{2012Icar..219..618G, 2023MNRAS.521.4927A}. 
However, it found that $k_{\rm net}$ is smaller than $k_{\rm rad}$ when $r_{\rm peb} \gtrsim 1~\si{mm}$ \citep[e.g.,][]{2017MNRAS.469S.755B, 2021MNRAS.508.4705B}.
As we assume $r_{\rm peb} = 1~\si{cm}$, we simply neglect the contribution of $k_{\rm net}$ and assume that $k_{\rm comet}$ is given by
\begin{equation}
k_{\rm comet} = k_{\rm rad}.
\end{equation}
Note that we still consider the thermal conductivity of pebbles ($k_{\rm peb}$) because of its contribution to the $k_{\rm rad}$ (see below and Appendix \ref{app:k_peb}).
For a pebble pile, the thermal conductivity due to radiation through the void space between pebbles, $k_{\rm rad}$, is given by
\begin{equation}
k_{\rm rad} = 8 \sigma_{\rm SB} T^{3} r_{\rm peb} {\mathscr F}_{\rm ex} f_{\rm k},
\label{eq:k_rad}
\end{equation}
where $r_{\rm peb} = 1~\si{cm}$ is the pebble radius, ${\mathscr F}_{\rm ex}$ is the radiative exchange factor, and $f_{\rm k}$ is the non-isothermal correction factor \citep[e.g.,][]{2022JGRE..12707191R}.
Heat diffusion and surface‐to‐surface radiation calculations by \citet{2022JGRE..12707191R} revealed that ${\mathscr F}_{\rm ex}$ is given by
\begin{equation}
{\mathscr F}_{\rm ex} = 0.739 + 0.629 {\left( \frac{1 - \phi_{\rm pack}}{\phi_{\rm pack}} \right)}^{1.031},
\end{equation}
and $f_{\rm k}$ is given by 
\begin{equation}
f_{\rm k} =  1.007 - 0.500 \arctan{\left[ 1.351 {\left( \frac{\phi_{\rm pack}}{\Lambda_{\rm peb}} \right)}^{0.741} \right]},
\label{eq:f_k}
\end{equation}
where
\begin{equation}
\Lambda_{\rm peb} = \frac{k_{\rm peb}}{8 \sigma_{\rm SB} r_{\rm peb} T^{3}}.
\end{equation}
The thermal conductivity of pebbles, $k_{\rm peb}$, is a function of $T$ and $C$ (see Appendix \ref{app:k_peb}).

\subsubsection{Specific heat capacity}

The specific heat capacity of comet, $c_{\rm comet}$, is given by
\begin{equation}
c_{\rm comet} = f_{\rm ice} c_{\rm ice} + f_{\rm rock} c_{\rm rock},
\end{equation}
where $c_{\rm ice}$ and $c_{\rm rock}$ are temperature-dependent specific heat capacities of ice and rock, respectively.
In this study, $c_{\rm ice}$ and $c_{\rm rock}$ are given by
\begin{equation}
c_{\rm ice} = {\left[ 90 + 749 {\left( \frac{T}{100~\si{K}} \right)} \right]}~\si{J.kg^{-1}.K^{-1}},
\label{eq:c_ice}
\end{equation}
and
\begin{equation}
c_{\rm rock} = 1043 {\left[ 1 - \exp{\left\{ - 0.35 {\left( \frac{T}{100~\si{K}} \right)} \right\}} \right]}~\si{J.kg^{-1}.K^{-1}},
\label{eq:c_rock}
\end{equation}
respectively \citep[see][]{2022MNRAS.514.3366M}.

\section{Results}
\label{sec:results}

We calculate the temperature profile of km-sized pebble-pile comets.
We regard $L_{\rm cry}$, $R_{\rm comet}$, and $t_{\rm acc}$ as parameters.
First, we present our fiducial case that uses $L_{\rm cry} = 9 \times 10^{4}~\si{J.kg^{-1}}$, $R_{\rm comet} = 2~\si{km}$, and $t_{\rm acc} = 5~\si{Myr}$ (see Section \ref{sec:exothermic}).
While we assume $L_{\rm cry} = 9 \times 10^{4}~\si{J.kg^{-1}}$ for the exothermic case \citep{1968JChPh..48..503G}, we use $- 9 \times 10^{4}~\si{J.kg^{-1}}$ for the endothermic case (see Section \ref{sec:endothermic}).
We set $R_{\rm comet} = 2~\si{km}$ as a typical radius of Jupiter-family comets \citep[e.g.,][]{2006Icar..182..527T, 2008ssbn.book..397L, 2011MNRAS.414..458S}.
As the smaller comets are more abundant in number, we explore the size dependence from $R_{\rm comet} = 0.4~\si{km}$ to $2~\si{km}$ in Section \ref{sec:depth}.
We also study the effect of $t_{\rm acc}$ ($= 4~\si{Myr}$ and $5~\si{Myr}$) in Section \ref{sec:depth}.
However, we fix some parameters (e.g., $\chi_{\rm rock}$, $r_{\rm peb}$, and $T_{\rm bg}$) and do not explore their effect. This is because our objective is demonstrating the effect of $L_{\rm cry}$ on the thermal history of km-sized comets in the early solar system.

\subsection{Exothermic case}
\label{sec:exothermic}

\begin{figure}
\begin{center}
\includegraphics[width=\columnwidth]{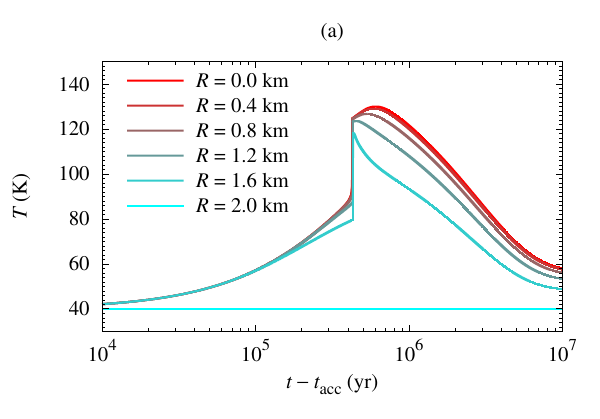}
\includegraphics[width=\columnwidth]{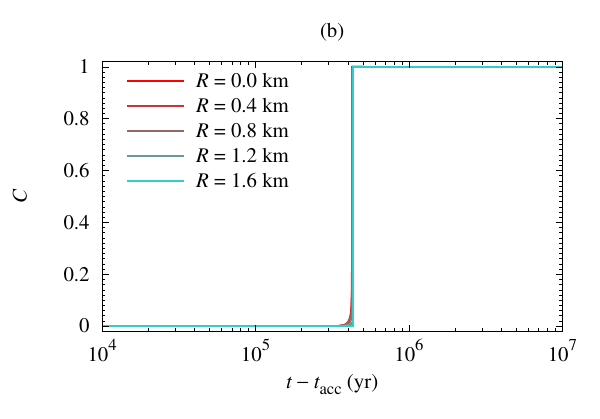}
\includegraphics[width=\columnwidth]{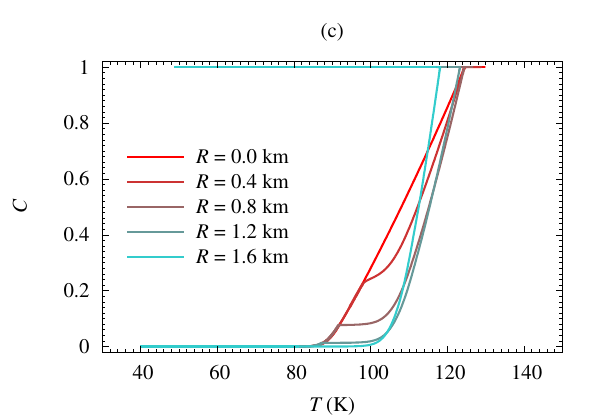}
\end{center}
\caption{
Thermal evolution history of a comet with $L_{\rm cry} = 9 \times 10^{4}~\si{J.kg^{-1}}$, $R_{\rm comet} = 2~\si{km}$, and $t_{\rm acc} = 5~\si{Myr}$.
Each line represents a different $R$ (see legend).
(a) Temporal evolution of the temperature of the cometary interior.
(b) Temporal evolution of the ice crystallinity.
(c) Evolutionary track of the ice crystallinity as a function of temperature.
In Panel (c), each track starts at the bottom left ($T = 40~\si{K}$ and $C = 0$).
}
\label{fig2}
\end{figure}

First, we present our fiducial thermal evolution history of a km-sized comet assuming the ice crystallization is exothermic.
Figure \ref{fig2}(a) illustrates the temporal evolution of the temperature of the cometary interior with $L_{\rm cry} = 9 \times 10^{4}~\si{J.kg^{-1}}$, $R_{\rm comet} = 2~\si{km}$, and $t_{\rm acc} = 5~\si{Myr}$.
In this calculation, the crystallization of amorphous ice is an exothermic process, and we can see a sharp increase of the temperature at $t - t_{\rm acc} \simeq 4.3 \times 10^{5}~\si{yr}$.
The temporal evolution of the temperature shown in Figure \ref{fig2}(a) is qualitatively consistent with that reported in earlier studies \citep[e.g.,][]{2022MNRAS.514.3366M}.

The radial profile of ice crystallinity also changes with time.
Figure \ref{fig2}(b) shows the temporal evolution of the ice crystallinity, $C$, for the exothermic case.
A sharp increase of $C$ is observed at $t - t_{\rm acc} \simeq 4.3 \times 10^{5}~\si{yr}$.
This corresponds to the sharp increase of the temperature in Figure \ref{fig2}(a).
The temperature during crystallization depends on the latent heat of crystallization.

\begin{figure}
\begin{center}
\includegraphics[width=\columnwidth]{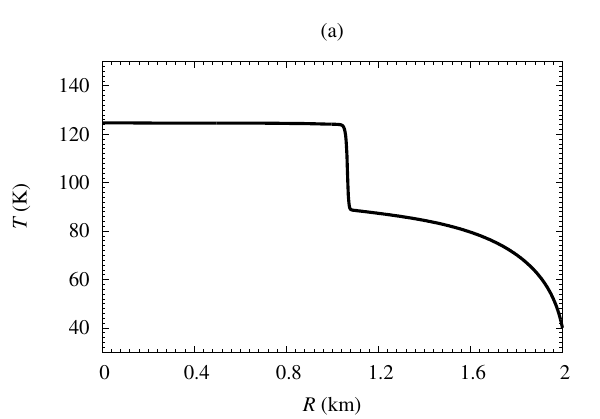}
\includegraphics[width=\columnwidth]{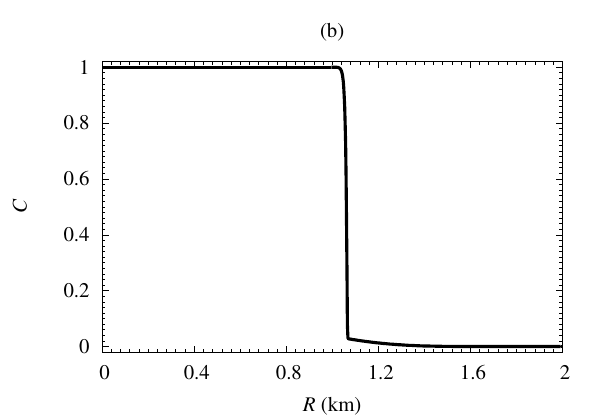}
\end{center}
\caption{
Radial distribution of (a) $T$ and (b) $C$ at $t = t_{\rm acc} + 4.294 \times 10^{5}~\si{yr}$.
Note that the parameters are the same as Figure \ref{fig2}.
}
\label{figX}
\end{figure}

Figure \ref{fig2}(c) represents the evolutionary track of the ice crystallinity as a function of temperature.
The temperature changes drastically during crystallization for the exothermic case.
The onset temperature for crystallization varies with the distance from the center, $R$, and the temperature at the end of crystallization also depends on $R$.
To explain these dependencies, we show the radial distribution of $T$ and $C$ at $t = t_{\rm acc} + 4.294 \times 10^{5}~\si{yr}$ (Figure \ref{figX}).
The temperature gradient is large at around the crystallization front ($R \simeq 1~\si{km}$), and it causes the large heat flux which is proportional to ${\partial}T / {\partial}R$.
At the same time, ice crystallization also triggers the increase of $T$, which is proportional to ${\rm d}C / {\rm d}t$.
As the balance between two processes varies with time ($t$) and position ($R$), the evolutionary tracks of $C$ depend on $R$ (see Figure \ref{fig2}(c)).

\subsection{Endothermic case}
\label{sec:endothermic}

\begin{figure}
\begin{center}
\includegraphics[width=\columnwidth]{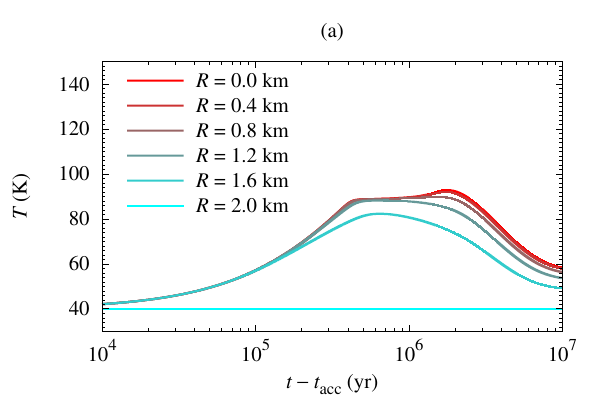}
\includegraphics[width=\columnwidth]{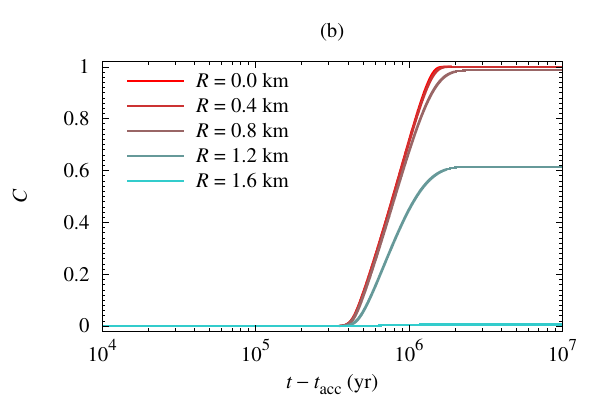}
\includegraphics[width=\columnwidth]{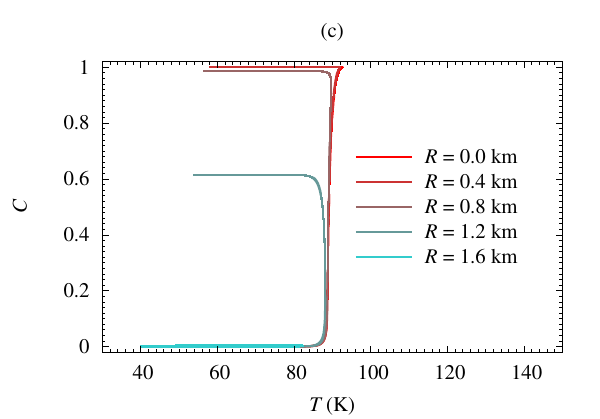}
\end{center}
\caption{
Same as Figure \ref{fig2}, but for $L_{\rm cry} = - 9 \times 10^{4}~\si{J.kg^{-1}}$.
}
\label{fig3}
\end{figure}

Next, we show the thermal evolution history of a km-sized comet assuming the crystallization of amorphous ice is an endothermic process.
Figure \ref{fig3}(a) shows the temporal evolution of the temperature of the cometary interior for $L_{\rm cry} = - 9 \times 10^{4}~\si{J.kg^{-1}}$.
Note that we set both $R_{\rm comet} = 2~\si{km}$ and $t_{\rm acc} = 5~\si{Myr}$, as in Section \ref{sec:exothermic}.
In this case, we can see a plateau of the temperature curve from $t - t_{\rm acc} \simeq 4 \times 10^{5}$--$1 \times 10^{6}~\si{yr}$.
This kind of plateau of the temperature curve is similar to the thermal evolution of carbonaceous chondrite parent bodies made of ice and rock \citep[e.g.,][]{2012NatCo...3..627F, 2013E&PSL.362..130F}.
This is because carbonaceous chondrite parent bodies experienced ice melting and aqueous alteration reactions, and the net heat generated via these two reactions is negative \citep[e.g.,][]{2011EP&S...63.1193W}.

The temporal evolution of the ice crystallinity also differs from that of the exothermic case.
Figure \ref{fig3}(b) represents the temporal evolution of $C$ for the endothermic case.
In this case, the crystallinity increases gradually with a timescale of several $10^{5}$ years.
In addition, Figure \ref{fig3}(c) indicates that the temperature is kept approximately constant during crystallization, and it barely depends on $R$ for the endothermic case.
This is because almost all heat generated by radioactivity is consumed for the crystallization of amorphous ice.

The final crystallinity at $R = 1.6~\si{km}$ clearly depends on whether ice crystallization is exothermic or endothermic.
For the exothermic case, the final crystallinity is $C \simeq 1$ at $R = 1.6~\si{km}$ (Figure \ref{fig2}(b)) while $C \simeq 0$ for the endothermic case (Figure \ref{fig3}(b)).
This suggests that the radial distribution of ice crystallinity at the end of the evolution phase highly depends on $L_{\rm cry}$ (see Section \ref{sec:radial}).

\subsection{Radial distribution of the crystallinity}
\label{sec:radial}

\begin{figure}
\begin{center}
\includegraphics[width=\columnwidth]{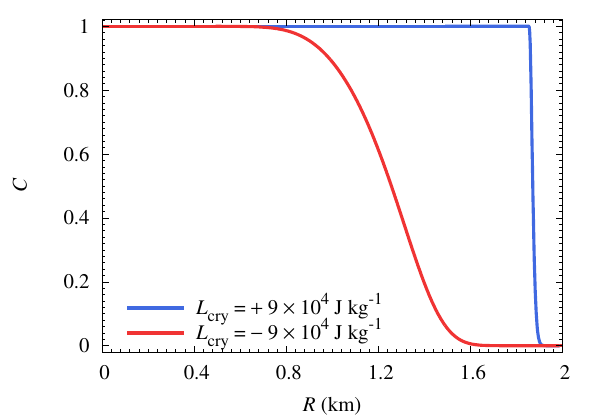}
\end{center}
\caption{
Radial distribution of the crystallinity at the end of calculation ($t = t_{\rm acc} + 10~\si{Myr}$).
}
\label{fig5}
\end{figure}

The radial distribution of ice crystallinity at the end of the evolution phase is of great interest.
The results at the end of the calculation ($t = t_{\rm acc} + 10~\si{Myr}$) for both exothermic and endothermic cases are summarized in Figure \ref{fig5}.
Though we find the similarities for both cases ($C \simeq 1$ at $R = 0~\si{km}$ and $C \simeq 0$ at $R = R_{\rm comet}$), there is a significant difference in $C$ from $R = 0.8~\si{km}$ to $R = 1.9~\si{km}$.
The region where the crystallinity is close to unity (i.e., the crystalline core in Figure \ref{fig1}(b)) for the exothermic case is wider than that for the endothermic case.
This result is intuitive as the release of latent heat for the exothermic case acts as a heat source and makes a wider region of the interior warm. 
In contrast, the consumption of heat for the endothermic case acts as a heat sink and makes the interior relatively cold.

\subsection{Depth of the amorphous ice mantle}
\label{sec:depth}

Finally, to study the survivability of amorphous ice in pebble-pile comets, we evaluate the depth of the amorphous ice mantle at the end of the evolution phase (Figure \ref{fig1}(b)). 
Here we define the depth of the amorphous ice mantle, $d_{x}$, as the distance from the surface where the ice crystallinity is lower than $x\%$ at $t = t_{\rm acc} + 10~\si{Myr}$.
We arbitrarily choose the threshold value, $x$, as $x = 10$ (i.e., $C = 0.1$) and $90$ (i.e., $C = 0.9$).
In our fiducial exothermic case ($L_{\rm cry} = 9 \times 10^{4}~\si{J.kg^{-1}}$, Section \ref{sec:exothermic}), both $d_{10}$ and $d_{90}$ are approximately $100~\si{m}$ ($d_{10} = 113~\si{m}$ and $d_{90} = 138~\si{m}$; see Figure \ref{fig5}).
In contrast, in our endothermic case ($L_{\rm cry} = - 9 \times 10^{4}~\si{J.kg^{-1}}$, Section \ref{sec:endothermic}), $d_{10} = 541~\si{m}$ and $d_{90} = 1013~\si{m}$ are significantly larger than those for the exothermic case (see Figure \ref{fig5}).
This clearly indicates that $L_{\rm cry}$ determines the depth $d_{x}$.

\begin{figure*}
\begin{center}
\includegraphics[width=1.5\columnwidth]{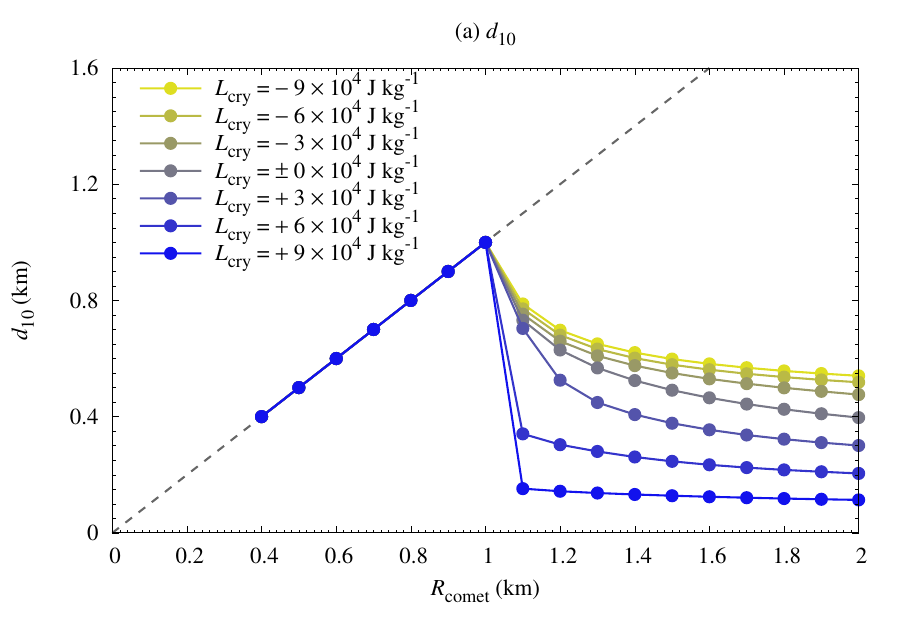}
\includegraphics[width=1.5\columnwidth]{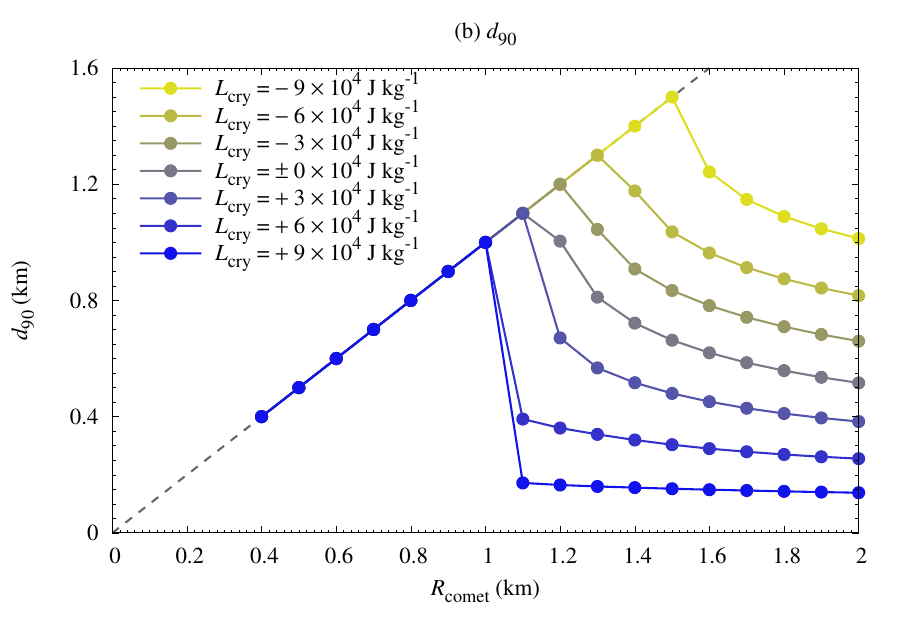}
\end{center}
\caption{
Depth of the amorphous ice mantle as a function of $R_{\rm comet}$. 
Each line depicts a different $L_{\rm cry}$ (see legend).
(a) For the threshold crystallinity of 10\% ($d_{10}$).
(b) For the threshold crystallinity of 90\% ($d_{90}$).
Note that we set $t_{\rm acc} = 5~\si{Myr}$ for all models shown here.
}
\label{fig7}
\end{figure*}

\begin{figure}
\begin{center}
\includegraphics[width=\columnwidth]{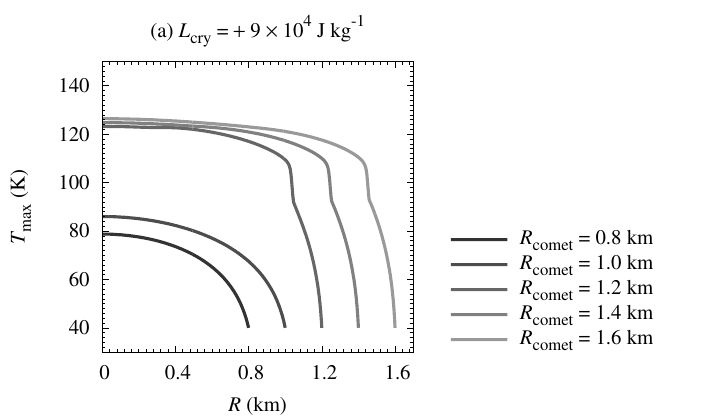}
\includegraphics[width=\columnwidth]{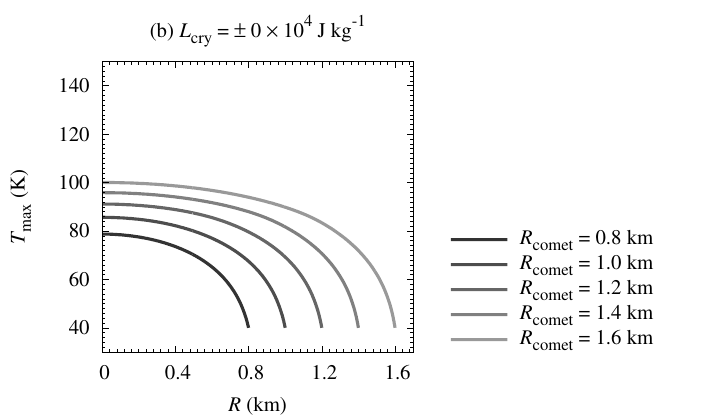}
\includegraphics[width=\columnwidth]{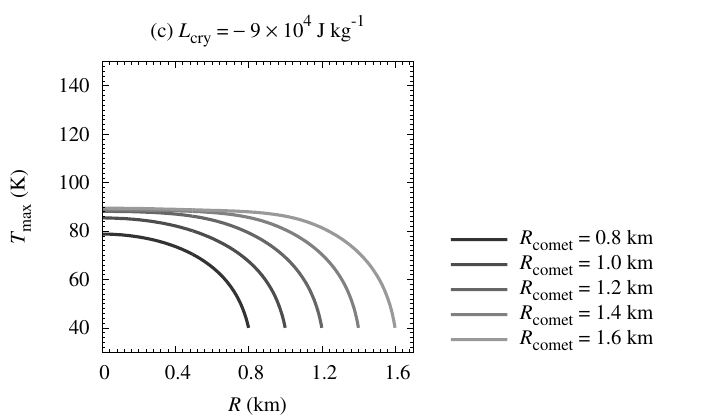}
\end{center}
\caption{
Radial distribution of the peak temperature during thermal evolution, $T_{\rm max}$.
Each line represents a different $R_{\rm comet}$ (see legend).
(a) For $L_{\rm cry} = 9 \times 10^{4}~\si{J.kg^{-1}}$.
(b) For $L_{\rm cry} = 0 \times 10^{4}~\si{J.kg^{-1}}$.
(c) For $L_{\rm cry} = - 9 \times 10^{4}~\si{J.kg^{-1}}$.
Note that we set $t_{\rm acc} = 5~\si{Myr}$ for all models shown here.
}
\label{fig:T_max}
\end{figure}

The depth of the amorphous ice mantle also depends on the radius of comets ($R_{\rm comet}$). 
Figure \ref{fig7} shows the depth of the amorphous ice mantle as a function of $R_{\rm comet}$. 
We also change $L_{\rm cry}$ but fix $t_{\rm acc} = 5~\si{Myr}$.
For $R_{\rm comet} \le 1~\si{km}$, the ice crystallinity kept below $10\%$ in the entire region regardless of the choice of $L_{\rm cry}$ (Figure \ref{fig7}(a)).
This can be also seen in the radial distribution of the peak temperature (Figure \ref{fig:T_max}).
In contrast, for $R_{\rm comet} > 1~\si{km}$, $d_{10}$ is always smaller than $R_{\rm comet}$; the crystallinity at the center is higher than $10\%$ regardless of $L_{\rm cry}$.
For the exothermic case of $L_{\rm cry} = 9 \times 10^{4}~\si{J.kg^{-1}}$, both $d_{10}$ and $d_{90}$ are nearly independent of $R_{\rm comet}$ when $R_{\rm comet} \ge 1.1~\si{km}$ (Figures \ref{fig7}(a) and (b)).
As $d_{90} \simeq d_{10}$ when $L_{\rm cry} \ge 6 \times 10^{4}~\si{J.kg^{-1}} ~\si{J.kg^{-1}}$, the crystallinity sharply changes at a certain depth (see also Figure \ref{fig5}).
In contrast, when $L_{\rm cry} \lesssim 0~\si{J.kg^{-1}}$, $d_{90}$ is significantly larger than $d_{10}$.
We also find that $d_{90}$ takes the maximum at $R_{\rm comet} > 1~\si{km}$, whereas $d_{10}$ takes the maximum at $R_{\rm comet} \simeq 1~\si{km}$.
Therefore, if $L_{\rm cry} \lesssim 0~\si{J.kg^{-1}}$, the ice in the deep interior might be still partially amorphous.

To understand the dependence of $d_{10}$ (and $d_{90}$) on $R_{\rm comet}$, we present the radial distribution of the peak temperature at each $R$, $T_{\rm max} {( R )}$, and its dependence on $R_{\rm comet}$.
Figure \ref{fig:T_max} illustrates the radial distribution of $T_{\rm max}$.
We find that the radial distribution of $T_{\rm max}$ is independent of $L_{\rm cry}$ when $R_{\rm comet} \le 1~\si{km}$, and $T_{\rm max} < 90~\si{K}$ in the entire region of the interior.
In other words,  $t_{\rm cry}$ is always longer than $1~\si{Myr}$ in the entire region when $R_{\rm comet} \le 1~\si{km}$ (see Figure \ref{fig:timescale} in Appendix \ref{app:timescale}).
This is similar to the previous findings that the radiogenic heating is less effective for a smaller comet \citep{2004come.book..359P}, and it is consistent with the fact that $C < 10\%$ in the entire region regardless of the choice of $L_{\rm cry}$.
In contrast, $T_{\rm max}$ varies significantly depending on $L_{\rm cry}$ when $R_{\rm comet} > 1~\si{km}$.
In the case of $L_{\rm cry} = 9 \times 10^{4}~\si{J.kg^{-1}}$, a sharp increase of $T_{\rm max}$ is observed at $R \approx R_{\rm comet} - d_{10}$.
In the cases of $L_{\rm cry} = 0 \times 10^{4}~\si{J.kg^{-1}}$, the sharp increase of $T_{\rm max}$ is absent.
In the case of $L_{\rm cry} = - 9 \times 10^{4}~\si{J.kg^{-1}}$, we see a plateau of $T_{\rm max}$ (see also Figure \ref{fig3}(a) and corresponding text in Section \ref{sec:endothermic}).

\begin{figure*}
\begin{center}
\includegraphics[width=1.5\columnwidth]{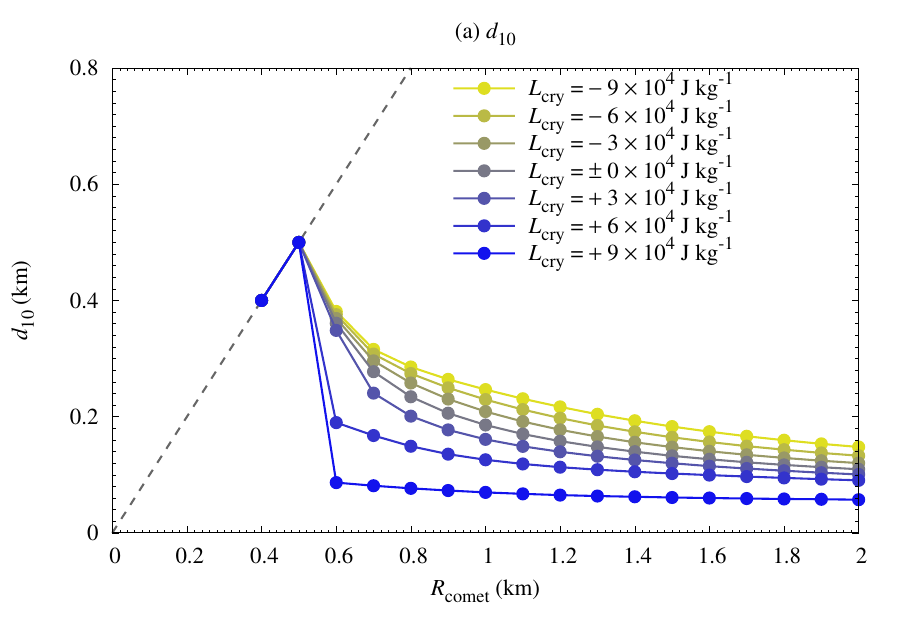}
\includegraphics[width=1.5\columnwidth]{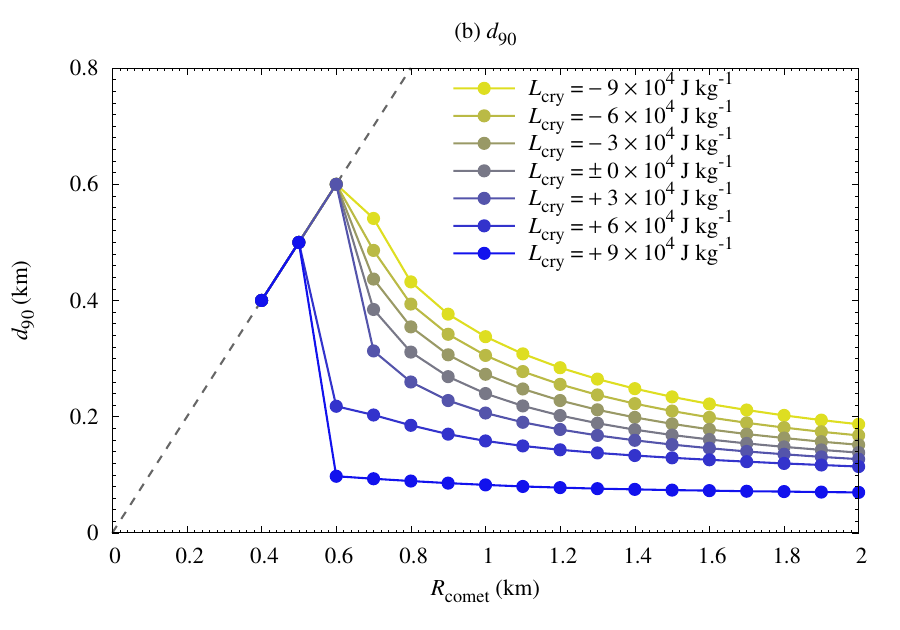}
\end{center}
\caption{
Same as Figure \ref{fig7}, but for $t_{\rm acc} = 4~\si{Myr}$.
}
\label{fig:4Myr}
\end{figure*}

The dependence of $L_{\rm cry}$ and $R_{\rm comet}$ on the depth of the amorphous mantle still holds in case of other $t_{\rm acc}$.
We set $t_{\rm acc} = 5~\si{Myr}$ in our fiducial model, but an earlier formation of comets with $t_{\rm acc} < 5~\si{Myr}$ might be possible. 
Chondrule-like objects from comet 81P/Wild 2 obtained by the \textit{Stardust} mission indicated that their accretion time is $t_{\rm acc} \gtrsim 3~\si{Myr}$ \citep[e.g.,][]{2012ApJ...745L..19O, 2015E&PSL.410...54N}.
In addition, the oxygen isotopic compositions and Mg\# (i.e., molar ratio of {MgO}/{(MgO + FeO)} in a mineral) of cometary chondrule-like objects are similar to those of CR chondrites \citep[e.g.,][]{2015E&PSL.410...54N, 2017E&PSL.465..145D}.
As the accretion age of CR chondrite parent body is estimated to be $t_{\rm acc} \gtrsim 4~\si{Myr}$ \citep[e.g.,][]{2017GeCoA.201..275S}, we can expect that the accretion age of comets may also be $t_{\rm acc} \gtrsim 4~\si{Myr}$, and comets were formed near the region where CR chondrite parent body was formed.
Therefore, we investigate the case of $t_{\rm acc} = 4~\si{Myr}$ (see Figure \ref{fig:4Myr}).

Although the transition value of $R_{\rm comet}$ for the depth of the amorphous ice mantle ($R_{\rm comet} \simeq 0.5~\si{km}$) is different from that for $t_{\rm acc} = 5~\si{Myr}$ ($R_{\rm comet} \simeq 1~\si{km}$), we can still observe the sharp change of $d_{10}$ and $d_{90}$.
For $L_{\rm cry} \ge 6 \times 10^{4}~\si{J.kg^{-1}} ~\si{J.kg^{-1}}$, both $d_{10}$ and $d_{90}$ are nearly independent of $R_{\rm comet}$ when $R_{\rm comet}$ is larger than the transition value.
Therefore, general trends described in this section would be common for other $t_{\rm acc}$.

\section{Discussion}
\label{sec:discussion}

\subsection{Survival time of amorphous mantle in the active phase}

We confirm that an amorphous mantle and a crystalline core would be formed after the thermal evolution in the evolution phase.
Our results also show that the depth of the amorphous mantle significantly depends on $L_{\rm cry}$ and $R_{\rm comet}$.
Here we evaluate the typical survival time of amorphous mantle in the active phase by an order-of-magnitude estimate.

During the active phase, the radius of a comet decreases by the activity-driven mass loss.
The mass loss rate is a strong function of the orbital position, and the total mass loss per one orbit, ${\Delta {M}_{\rm loss}}$, has been evaluated for several comets by space missions \citep[e.g.,][]{2020SSRv..216...44C} and ground-based observations \citep[e.g.,][]{1999AJ....117.1056J}.
The change of the radius in one orbital period, ${\Delta R_{\rm comet}}$, is given by
\begin{equation}
{\Delta R_{\rm comet}} \simeq - \frac{\Delta {M}_{\rm loss}}{4 \pi {R_{\rm comet}}^{2} \rho_{\rm comet}}.
\end{equation}
For comet 67P/Churyumov--Gerasimenko, \citet{2019MNRAS.483.2337P} reported that ${\Delta {M}_{\rm loss}} = {( 1.05 \pm 0.34 )} \times 10^{10}~\si{kg}$.
Assuming $R_{\rm comet} \sim 2~\si{km}$ for comet 67P, we obtain ${\Delta R_{\rm comet}} \sim 0.4~\si{m}$.
\citet{2015A&A...583A..38B} also reported that ${\Delta R_{\rm comet}} \sim {( 1 \pm 0.5 )}~\si{m}$ and it is not far from our order-of-magnitude estimate.
The orbital period of comet 67P is $P_{\rm orb} = 6.45~\si{yr}$.
Then the survival time of amorphous mantle, $t_{\rm survival}$, could be evaluated as follows:
\begin{eqnarray}
t_{\rm survival} & = & \frac{d_{\rm amo}}{\Delta R_{\rm comet}} P_{\rm orb}, \nonumber \\
                 & = & 1.6 \times 10^{3} {\left( \frac{d_{\rm amo}}{100~\si{m}} \right)} {\left( \frac{\Delta R_{\rm comet}}{0.4~\si{m}} \right)}^{-1} {\left( \frac{P_{\rm orb}}{6.45~\si{yr}} \right)}~\si{yr},
\end{eqnarray}
where $d_{\rm amo}$ is the depth of the amorphous mantle at the beginning of the active phase.
Our estimate indicates that the amorphous ice mantle of $d_{\rm amo} \sim 100~\si{m}$ on a short-period comet would be removed by the activity-driven mass loss within several thousand years.
The $t_{\rm survival}$ might be comparable to or shorter than the dynamical lifetime of Jupiter-family comets \citep[$\sim 10~\si{kyr}$;][]{2017A&A...598A.110R}, and the removal of amorphous mantle could proceed in the active phase.

We are aware that our estimates depend on the choice of the reference comet.
For example, the radius of comet 103P/Hartley 2 is $R_{\rm comet} = {( 0.57 \pm 0.08 )}~\si{km}$ \citep{2009PASP..121..968L}.
The total mass loss per one orbit is ${\Delta {M}_{\rm loss}} \sim 1.2 \times 10^{10}~\si{kg}$ for the 1997 apparition and ${\Delta {M}_{\rm loss}} \sim 4 \times 10^{9}~\si{kg}$ for the 2010 apparition if we assume that the dust-to-gas ratio is $\sim 1$ \citep{2013Icar..222..550T}.
Then ${\Delta R_{\rm comet}} \sim 4~\si{m}$ for the 1997 apparition when $\rho_{\rm comet}$ is close to that of comet 67P.
To quantify the impact of activity-driven mass loss on the survivability of amorphous ice mantle, future observations on evaluating the variation of ${\Delta R_{\rm comet}}$ would be necessary.

\subsection{Size frequency distribution of comets}

The presence of an amorphous mantle may relate to the size and activity of comets. 
Several observations studied the size-frequency distribution of comets \citep[e.g.,][]{2006Icar..182..527T, 2011MNRAS.414..458S, 2016A&A...585A...9L}.
\citet{2006Icar..182..527T} found a depletion at sizes smaller than $R_{\rm comet} \sim 1.5~\si{km}$ for the cumulative size distribution of Jupiter-family comets \citep[see also][]{2011MNRAS.414..458S}.
They also reported that fractions of active surface area ($f_{\rm active}$) decreases with increasing $R_{\rm comet}$, and $f_{\rm active} \lesssim 0.01$ for comets with $R_{\rm comet} \gtrsim 3~\si{km}$.
These trends might be closely tied with some physical processes which cause the rapid formation of a thermally-insulating dust crust on larger nuclei.
In Section \ref{sec:depth}, we found that the depth of the amorphous ice mantle depends on $R_{\rm comet}$ at the end of the evolution phase.
When the crystallization process is an exothermic reaction with $L_{\rm cry} \ge 6 \times 10^{4}~\si{J.kg^{-1}}$, there is a sharp transition for the dependence of $d_{10}$ (or $d_{90}$) on $R_{\rm comet}$ (see Figure \ref{fig7}).
As the process of phase transition from amorphous ice to crystalline ice can trigger cometary activity \citep[e.g.,][]{2009Icar..201..719M}, the activity might indicate the presence of amorphous ice.
The weak activity on large comets might be associated with the low abundance of amorphous ice in the subsurface or the small thickness of the amorphous ice mantle.
Note that efficient sintering of ice could also explain the weak activity of larger comets.
The efficiency of pressure sintering is high in large comets, and sintering would act at higher temperatures than crystallization, judging from porosity evolution modeling for both rocky \citep[e.g.,][]{2014A&A...567A.120N, 2021Icar..35814166N} and icy objects \citep[e.g.,][]{2019Icar..326...10B, 2022Icar..37314776B, 2020A&A...633A.117N}.
The material strength of nucleus would increase as sintering progresses. 
If future space missions could conduct some artificial impacts on inactive comets as a similar way of Hayabusa 2 \citep{2020Sci...368...67A} or DART \citep{2023Natur.616..443D}, it would help us to understand the fate of icy objects.
We also hope future observations on comets' size will shed light on the effect of the presence/absence of amorphous ice mantle on the formation of dust crust during the active phase.

\subsection{Comets as rubble piles formed via collisional disruptions}

Comets are usually regarded as survivors of primordial planetesimals as they are rich in supervolatiles and have a higher porosity.
However, several studies have investigated the possibility of comet formation through collisional disruptions of larger bodies \citep[e.g.,][]{2017A&A...597A..62J, 2018NatAs...2..379S, 2020Icar..35013867J}.
\citet{2018NatAs...2..379S} reported that km-sized elongated or bilobate comets could be formed via catastrophic collisional disruptions and subsequent gravitational reaccumulation.
\citet{2017A&A...597A..62J} revealed that sub-catastrophic disruptions could also form km-sized elongated or bilobate comets.
These studies also found that these comets would maintain volatiles and high porosity throughout the events when the collision velocity is a few $\si{km.s^{-1}}$ or lower.
\citet{2023PSJ.....4....4S} investigated the survivability of amorphous ice during collision events. 
They reported that amorphous ice could survive when the collision velocity is lower than $\sim 2~\si{km.s^{-1}}$, although it depends on the size and thermophysical properties of impactors.
When the rubble-pile comets have experienced those collisions, the layered structure (i.e., amorphous mantle and crystalline core) may be disrupted.

We note, however, that large icy objects might be thermally altered before disruption if those 10--100 km-sized bodies formed at around $t_{\rm acc} = 5~\si{Myr}$ or earlier \citep{2022MNRAS.514.3366M}.
For the rubble-pile hypothesis, materials originating from the thermally altered interior could be found both in the interior and on the surface of rubble-pile comets.
However, cometary silicate grains are usually anhydrous and they have never experienced aqueous alteration \citep[e.g.,][]{2006Sci...314.1735Z}, and parent bodies of comets should avoid thermal processes (e.g., aqueous alteration and thermal metamorphism).
One plausible solution is that comets' parent bodies were formed in the solar nebula later than 5 Myrs after CAI formation \citep[e.g.,][]{2019ApJ...875...30N}.
Another solution might be that the radioactive nuclide ${}^{26}{\rm Al}$ was heterogeneously distributed in the solar nebula and comets would be formed in ${}^{26}{\rm Al}$-poor regions \citep[e.g.,][]{1995Metic..30..365M, 2011ApJ...735L..37L, 2022MNRAS.514.3366M}.
Future studies on the coupled thermal--collisional evolution of large objects in the outer solar system would be the key to unveiling whether comets are pebble piles or rubble piles.
Understanding of the origin of the radioactive nuclide in the early solar system is also essential \citep[e.g.,][]{2022arXiv220311169D, 2023A&A...670A.105A, 2023ApJ...947L..29A, 2023arXiv231201948S}.

\subsection{Caveats in our current model}
\label{sec:caveat}

We would like to describe some caveats in this study.
We have made some assumptions to use our current one-dimensional numerical codes. 
While the actual shape of comets is non-spherical \citep[e.g.,][]{2017A&A...607L...1P}, we assume a spherical comet nucleus as well as previous studies \citep[e.g.,][]{2004come.book..317M, 2004come.book..359P}. 
To model a specific comet (e.g., 67P/Churyumov--Gerasimenko), we might need to consider updating our code to the three-dimensional code.
However, our current focus is evaluating the effect of latent heat of crystallization at the earlier evolution of comets.
Since we have no idea of initial comets' shape, our assumption of spherical nuclei would be valid.

Assuming the constant $R_{\rm comet}$ for simplicity, we neglect the temporal change of $R_{\rm comet}$ in the evolution phase.
The size evolution of comets would be caused by mass accretion or erosion/destruction \citep[e.g.,][]{2021MNRAS.505.5654D}.
These processes are closely related to the formation and dynamical evolution of small icy bodies \citep[e.g.,][]{2022arXiv220600010K, 2022arXiv221204509S}.
The sintering of porous icy aggregates would also cause the shrinking of comets \citep[e.g.,][]{2021MNRAS.505.5654D, 2022MNRAS.514.3366M}.
The shrinking of comets in the evolution phase would suppress the heating of cometary interior, and thus the size of the crystalline core might become larger than our results.
The impacts of size change on the early thermal evolution of comets should be quantitatively investigated in future studies.

Regardless of gas species trapped within the impure amorphous ice, they would be released upon the crystallization \citep{2001GeoRL..28..827K}.
While we focus on investigating the importance of the latent heat of impure amorphous ice, released gas would also transfer the heat and mass in reality \citep[e.g.,][]{1995Icar..117..420P, 1999AdSpR..23.1299S, 2001AJ....121.2792D, 2004come.book..359P, 2023MNRAS.522.2081P}.
However, we ignore the effects of released gas upon the crystallization for simplicity \citep[e.g.,][]{1993MNRAS.260..819Y, 1993JGR....9815079H, 1997P&SS...45..827S, 2017ApJ...842...11S}.
This is because the amount of trapped gas within the amorphous ice is small (up to a few \% in the endothermic case and much less than a few \% in the exothermic case).
Thus, heat transported by advection of released gas would be negligible \citep{1999AdSpR..23.1299S}, and we do not consider the effect of released gas upon the crystallization on the heat transfer in our current model.
To assess the effects of released gas, we need to solve the equations of mass and energy conservation simultaneously \citep[e.g.,][]{2023MNRAS.522.2081P} which remains a task to be explored in future studies.

\section{Conclusions}
\label{sec:conclusion}

Comets are small icy objects believed to have formed in the outer region of the solar nebula, and cometary ice is thought to be amorphous rather than crystalline at the epoch of their accretion \citep[e.g.,][]{2008SSRv..138..147P, 2022arXiv220905907P}.
The release of the latent heat of ice crystallization could be the pivotal process for the thermal evolution of comets.
However, the value of the latent heat for the crystallization of cometary ice strongly depends on the abundance of impurities in ice \citep{2001GeoRL..28..827K}.

We performed one-dimensional simulations of the thermal evolution of km-sized comets and investigated the dependence on the latent heat of ice crystallization due to impurities in ice.
We confirmed that the runaway crystallization occurs when the latent heat is positive (Figure \ref{fig2}(b)) as reported in earlier studies \citep[e.g.,][]{2022MNRAS.514.3366M}.
In contrast, in the negative latent heat case representing the impure ice, there is no runaway crystallization (Figure \ref{fig3}(b)).
We also found that the depth where amorphous ice can survive significantly depends on the latent heat of ice crystallization (Figures \ref{fig5} and \ref{fig7}).
When $R_{\rm comet} > 1~\si{km}$ and the cometary ice have the positive latent heat ($L_{\rm cry} = + 9 \times 10^{4}~\si{J.kg^{-1}}$), the depth of the amorphous ice mantle becomes approximately $100~\si{m}$. 
On the other hand, the depth of the amorphous ice mantle becomes larger than several hundred meters when $L_{\rm cry}$ is negative.
Our results suggest that the spatial distribution of the crystallinity of ice in a comet nucleus might be different from the previous work.
This is because the latent heat due to the pure/impure water ice governs the presence/absence of the runaway crystallization.
Our results also showed that the depth where amorphous ice can survive depends on the accretion age of comets (Figure \ref{fig:4Myr}).
As we fix some properties of comets (e.g., the volume fraction of ice and the thermophysical properties), we expect that such parameters may also affect the depth.
Nevertheless, the general trends described in this study would hold.
Future studies on material properties of comets and their formation processes would be essential for a more quantitative discussion on their thermal history.

Since the crystallization process of impure ice is still under debate \citep[e.g.,][]{2021ApJ...918...45K}, we expect that future experiments will provide some evidence on this issue. 
Future comet-related missions will also help us to better understand the physicochemical environment of the early solar system; the comet sample return missions will be the best that provides a new Rosetta Stone that we have never seen \citep[e.g.,][]{2018LPI....49.1332S, 2023wakita}.

\begin{ack}
We are grateful to editors Takayuki Muto and Masaki Ando for their helpful comments and kind handling.
The authors wish to express their cordial thanks to the referee Wladimir Neumann for constructive comments.
The authors also thank Hitoshi Miura and Sin-iti Sirono for fruitful discussions and comments.
\end{ack}

%






{\appendix

\section{Timescales for heat conduction to the next mesh and for crystallization}
\label{app:timescale}

We describe the temperature dependence of $t_{\rm mesh}$ and $t_{\rm cry} + t_{\rm cry, min}$ in Figure \ref{fig:timescale}.
We note that $t_{\rm mesh}$ takes the maximum when $C = 0$ and it takes the minimum when $C = 1$.
In our simulations, $1~\si{yr} \lesssim t_{\rm mesh} \lesssim 10~\si{yr}$ and it weakly depends on $T$.
When we set $t_{\rm cry, min} = 10^{9}~\si{s}$, $t_{\rm cry} + t_{\rm cry, min}$ is always larger than $t_{\rm mesh}$.

Here we note that an unphysical increase of the temperature is triggered in numerical simulations without $t_{\rm cry, min}$.
When we do not consider $t_{\rm cry, min}$, we see an artificial peak of the temperature curve at $t - t_{\rm acc} \approx 0.43~\si{Myr}$ in Figure \ref{fig:no_min}, which illustrates the temporal evolution of the temperature of the cometary interior with $L_{\rm cry} = 9 \times 10^{4}~\si{J.kg^{-1}}$, $R_{\rm comet} = 2~\si{km}$, and $t_{\rm acc} = 5~\si{Myr}$ (cf.~Figure \ref{fig2}(a)).
To avoid such an unphysical increase, we need to take into consideration a moderate value of $t_{\rm cry, min}$.

\begin{figure}
\begin{center}
\includegraphics[width=\columnwidth]{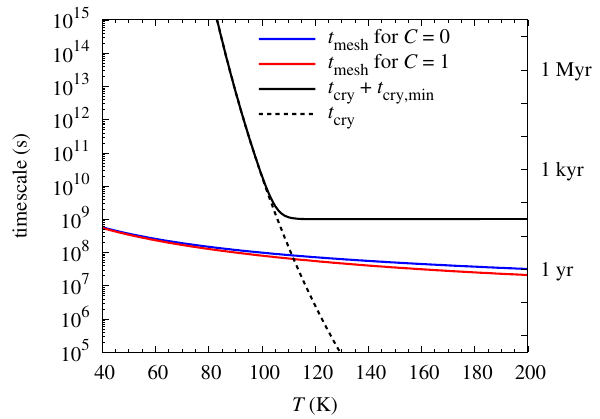}
\end{center}
\caption{
Temperature dependence of $t_{\rm mesh}$ and $t_{\rm cry} + t_{\rm cry, min}$.
}
\label{fig:timescale}
\end{figure}

\begin{figure}
\begin{center}
\includegraphics[width=\columnwidth]{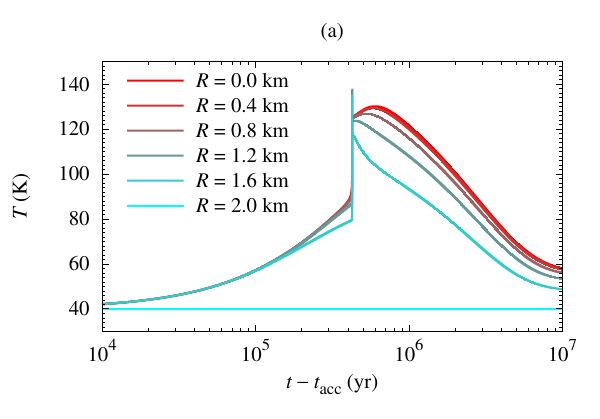}
\includegraphics[width=\columnwidth]{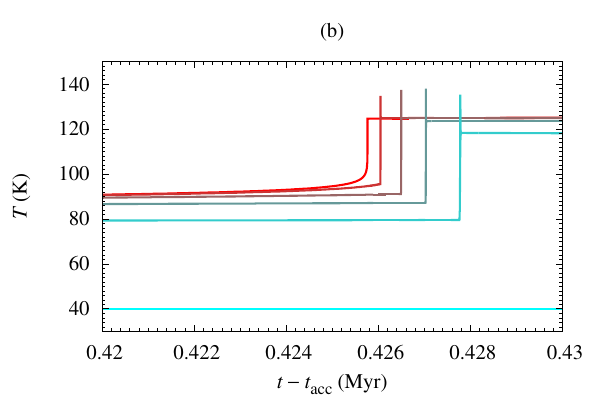}
\end{center}
\caption{
Temporal evolution of the temperature of the cometary interior with $L_{\rm cry} = 9 \times 10^{4}~\si{J.kg^{-1}}$, $R_{\rm comet} = 2~\si{km}$, and $t_{\rm acc} = 5~\si{Myr}$ (cf.~Figure \ref{fig2}(a)).
Here we do not consider $t_{\rm cry, min}$, and we can see an artificial peak of the temperature curve.
(a) For the entire period of thermal evolution.
(b) For zooming-in on the onset of crystallization.
}
\label{fig:no_min}
\end{figure}

\section{Thermal conductivity of pebbles}
\label{app:k_peb}

When pebbles are aggregates consisting of submicron-sized icy particles, the thermal conductivity of pebbles, $k_{\rm peb}$, is given as follows \citep[e.g.,][]{2012Icar..219..618G, 2023MNRAS.521.4927A}:
\begin{equation}
k_{\rm peb} = 2 k_{\rm mat} \frac{r_{\rm c, par}}{r_{\rm par}} f_{\rm peb},
\label{eq:k_peb}
\end{equation}
where $k_{\rm mat}$ is the material thermal conductivity, $r_{\rm par}$ is the particle radius, $r_{\rm c, par}$ is the interparticle contact radius, and $f_{\rm peb}$ is a dimensionless function associated with packing geometry.
The ratio of $r_{\rm c, par}$ to $r_{\rm par}$ is given by the following equation \citep[e.g.,][]{1971RSPSA.324..301J}:
\begin{equation}
\frac{r_{\rm c, par}}{r_{\rm par}} = {\left[ \frac{9 \pi \gamma_{\rm par} {\left( 1 - {\nu_{\rm par}}^{2} \right)}}{2 E_{\rm par} r_{\rm par}} \right]}^{1/3}.
\end{equation}
We ignore the change of $r_{\rm c, par}$ due to sintering for simplicity \citep[e.g.,][]{2017ApJ...842...11S, 2019A&A...628A..77G, 2021Icar..35814166N}.
\citet{2019Icar..324....8A} found that $f_{\rm peb}$ is given as follows:
\begin{equation}
f_{\rm peb} = 0.784 {\phi_{\rm peb}}^{1.99} {Z_{\rm peb}}^{0.556},
\end{equation}
where $Z_{\rm peb}$ denotes the average coordination number that also depends on $\phi_{\rm peb}$ \citep{2019PTEP.2019i3E02A, 2019Icar..324....8A}:
\begin{equation}
Z_{\rm peb} = 2 + 9.38 {\phi_{\rm peb}}^{1.62}.
\end{equation}

In this study, we assume that each submicron-sized icy particle is covered with an ice mantle and the heat conduction through the interparticle contact depends on the material thermal conductivity of the ice mantle \citep[e.g.,][]{1998A&A...330..375G, 2020MNRAS.497.1166A}.
It is known that $k_{\rm mat}$ depends on whether ice is amorphous or crystalline \citep[e.g.,][]{1980Sci...209..271K, 1992ApJ...388L..73K, 2002PhRvB..65n0201A}.
For amorphous ice, material thermal conductivity is given by
\begin{equation}
k_{\rm amo} = {\left[ 2.348 \times 10^{-1} {\left( \frac{T}{100~\si{K}} \right)} + 2.82 \times 10^{-2} \right]}~\si{W.m^{-1}.K^{-1}},
\label{eq:k_mat_amo}
\end{equation}
while for crystalline ice, it is given by
\begin{equation}
k_{\rm cry} = 5.67 {\left( \frac{T}{100~\si{K}} \right)}^{-1}~\si{W.m^{-1}.K^{-1}}.
\label{eq:k_mat_cry}
\end{equation}
Both $k_{\rm amo}$ and $k_{\rm cry}$ are set to be identical to those assumed in \citet{2022MNRAS.514.3366M}.
In this study, we consider that the ice mantle is a mixture of amorphous and crystalline ices, and $k_{\rm mat}$ is given by a simple function of $C$ as follows:
\begin{equation}
k_{\rm mat} = {\left( 1 - C \right)} k_{\rm amo} + C k_{\rm cry}.
\end{equation}

}

\bibliography{example}{}
\bibliographystyle{aasjournal}

\end{document}